# *In Situ* Imaging of the Conducting Filament in a Silicon Oxide Resistive Switch

Jun Yao[1], Lin Zhong[2,3*], Douglas Natelson[2,4*], James M. Tour[3,5*]

[1]Applied Physics Program through the Department of Bioengineering; [2]Department of Electrical and Computer Engineering; [3]Department of Computer Science; [4]Department of Physics and Astronomy; [5]Departments of Chemistry and Mechanical Engineering and Materials Science, Rice University, 6100 Main St., Houston, Texas 77005

*To whom correspondence should be addressed. Emails: lzhong@rice.edu; natelson@rice.edu; tour@rice.edu

The nature of the conducting filaments in many resistive switching systems has been elusive. Through *in situ* transmission electron microscopy, we image the real-time formation and evolution of the filament in a silicon oxide resistive switch. The electroforming process is revealed to involve the local enrichment of silicon from the silicon oxide matrix. Semi-metallic silicon nanocrystals with structural variations from the conventional diamond cubic form of silicon are observed, which likely accounts for the conduction in the filament. The growth and shrinkage of the silicon nanocrystals in response to different electrical stimuli show energetically viable transition processes in the silicon forms, offering evidence to the switching mechanism. The study here also provides insights into the electrical breakdown process in silicon oxide layers, which are ubiquitous in a host of electronic devices.

The understanding of the mechanisms in resistive switching materials is essential for their applications in future nonvolatile memory and logic devices (*1-3*). When in the ON state, conduction in many such systems takes place through a filament linking the source and drain electrodes, rather than throughout the channel material. While the confinement of the filament in many resistive switching materials is considered advantageous for further device scaling (*4*), it also presents considerable challenges in mechanistic probing. Techniques such as surface conductance mapping provide useful topological information regarding the filaments (*5-7*), yet the embedded nature of the filament makes further information, such as materials composition, difficult to obtain. Recently, by using transmission electron microscopy (TEM), more structural and compositional details in the filaments were revealed (*8-10*). Nevertheless, the isolation of the filament from the surrounding solid-electrolyte matrix, usually accomplished by cutting with a focused ion beam (*8-10*), is destructive and potentially alters the structural and compositional completeness of the filament. Furthermore, information such as the formation and evolution of the filament essential to the switching mechanism is largely missing in the *ex situ* imaging. Here we demonstrate the *in situ* imaging of the semi-metallic Si filament in a silicon oxide ($SiO_x$, $x \sim 2$) resistive switch, and this could have implications in a wide range of silicon oxide breakdown phenomena.

$SiO_x$ resistive switching memory is attractive for its fully CMOS (complementary metal-oxide-semiconductor) compatible material composition and processing (*4*). Two types of resistive switching memory have been constructed based on $SiO_x$. The first type belongs to the programmable metallization cell (*4*), which relies on the formation and rupture of the metal filament injected from the electrode (*11*). It features bipolar switching behavior and is extrinsic to $SiO_x$ with $SiO_x$ merely serving as the passive solid electrolyte. The second type shows unipolar current-voltage (*I-V*) characteristics with the reset voltage larger than the set value (Fig. 1A). The filamentary switching was recently revealed to be the intrinsic property of $SiO_x$ as it is electrode-independent (*9, 12*). The study here focuses on this type of intrinsic resistive switching in $SiO_x$. While *ex situ* imaging revealed a silicon-rich switching site (*9*), information regarding conduction and switching mechanisms was largely missing. Since soft breakdown is involved in the electroforming process (*1*), the study here could also provide insights into the general electrical breakdown process in $SiO_x$, a material that plays a ubiquitous role in semiconductor electronics.

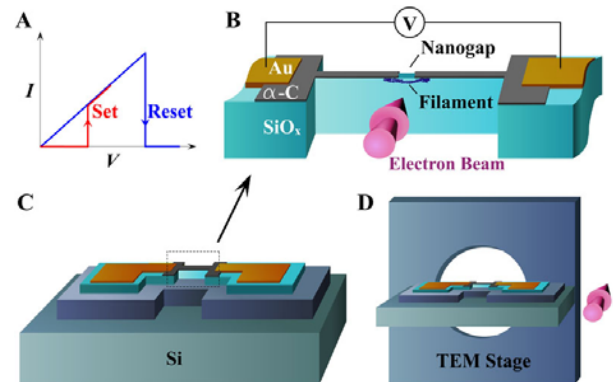

**Fig. 1.** *I-V* behaviors and general device embodiments. (**A**) Schematic of the unipolar switching in $SiO_x$, showing the typical set (red) and reset (blue) *I-V* curves (*9*). (**B**) Schematic of the $SiO_x$ thin-walled structure for *in situ* TEM imaging. The switching region in $SiO_x$ is localized by the nanogap generated in the α-C layer and imaged by TEM. (**C** and **D**) Schematics of the multi-stage structure and its arrangement with respect to the TEM stage. The pink arrows indicate the TEM electron beam imaging direction.

Figs. 1B-D shows the schematics of the device design and setup for the *in situ* imaging. The imaging region consists of a $SiO_x$ thin-walled structure covered by a layer of amorphous carbon (α-C), which is connected to external electrical inputs (Fig. 1B). By electrical breakdown in the α-C layer, a disruption region or nanogap can be produced as we described previously in planar carbon-coated $SiO_x$ devices (*12*). The broken ends of α-C layer then serve as the electrodes for the $SiO_x$ in the nanogap region. The use of α-C as the electrode material eliminates possible extrinsic effects from metals (*11*), and the electrical breakdown-generated nanogap provides an easy method for the fabrication of closely spaced electrodes atop a thin-walled structure. The confinement

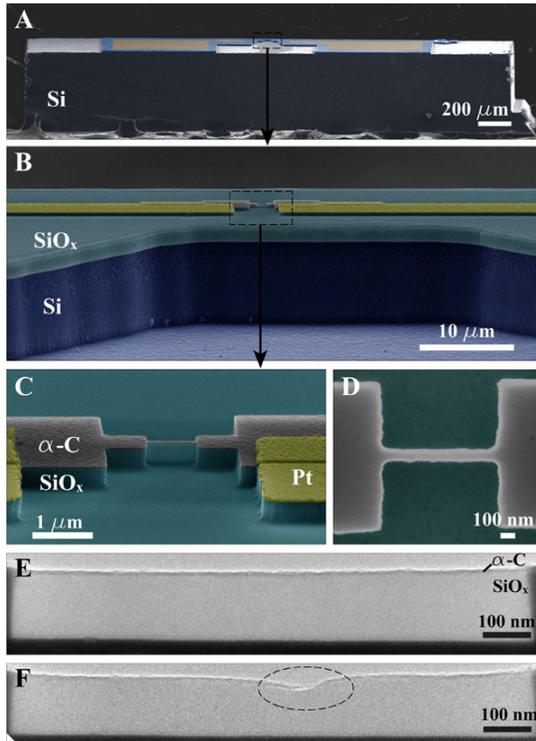

**Fig. 2.** SEM false-color images and TEM images of the electronic device. (**A-C**) Perspective SEM images of the multi-stage device for *in situ* imaging. (**D**) A top view of the $SiO_x$ thin-walled structure. (**E**) Low-magnification TEM image of the $SiO_x$ thin-walled structure with the α-C layer on top. (**F**) The same structure as in (E) with the α-C layer subjected to electrical breakdown. The circled region shows the nanogap in the α-C layer.

from the nanogap pre-localizes the switching site so that it can be constantly monitored from before the electroforming or filament formation and throughout the experiment. During imaging, the electron beam from the TEM system travels perpendicularly across the $SiO_x$ thin-walled structure. In this configuration, the $SiO_x$ and the α-C layer are spatially separated in the imaging plane, minimizing possible interference from the electrode material. Practically, for the successful passage of the electron beam across the nanogap region, a multi-stage design with the $SiO_x$ thin-walled structure is required (Fig. 1C). This multi-stage device is then vertically mounted on a home-built TEM stage that is capable of *in situ* electrical characterization (Fig. 1D).

Figs. 2A-D shows a series of scanning electron microscopy (SEM) images of the multi-stage device (see Fig. S1 for fabrication details). The thickness of the $SiO_x$ thin-walled structure is ~ 100 nm for the electron-beam transparency, and the length is ≤ 1 μm to reduce the resistance of the α-C electrodes (~ 20 nm thick). The successful imaging of the pristine structure by TEM is shown in Fig. 2E. After electrical breakdown in the α-C layer, a nanogap is generated in the α-C layer atop the $SiO_x$ thin-walled structure (Fig. 2F).

Fig. 3 shows a series of high-resolution TEM images of the nanogap region (right panels) with respect to different *I-V* responses (left panels). Note that during the electrical characterization, the electron beam was temporally blocked to exclude beam impact (*13*). Immediately after the electrical breakdown in the α-C layer, a nanogap of ~ 15 nm is formed (Fig. 4A). Both the $SiO_x$ at the nanogap region and the $SiO_x$ far from the nanogap show amorphous silica features. Because of the disruption in the α-C layer, the device shows little conduction during the subsequent voltage sweep, until at ~ 12 V the current suddenly increases (light grey curve in the plot of Fig. 3B). This conductance increase features the beginning of the electroforming process in $SiO_x$ (*9*). The device is subsequently electroformed, showing the characteristic *I-V* curve featuring current increase (at ~ 5 V) and decrease (at ~ 10 V) that define the typical set and reset processes, respectively (grey curve). The device is set to the low-resistance (ON) state (red curve). The immediate TEM imaging shows morphological changes at the nanogap region (Fig. 3B), as is often associated with the electroforming process (*8, 9, 14*). Specifically, out of the amorphous background, ~ 3 nm regions of nanocrystalline structure (based on the appearance of lattice fringes) appear at the nanogap (bounded region and inset in Fig. 3B). The apparent lattice spacing of the nanocrystal based on the fringes is distinct from that of the α-C, indicating a different material form.

Due to the beam impact (see below), the formed ON and switching states degrade after imaging (light grey curve in Fig. 3C). A subsequent re-electroforming process is involved to set the device back to ON (grey and red curves). The immediate imaging shows growth in the nanocrystal (bounded region and inset in Fig. 3C). The partially degraded ON state after this imaging is compensated by a set process (red curve in Fig. 3D). For the subsequent voltage sweep to 14 V, a sudden decrease in the conductance occurs at ~ 12 V, featuring the typical reset process (*9*). The immediate imaging shows prominent shrinkage in the size of the nanocrystal (bounded region and inset in Fig. 4D).

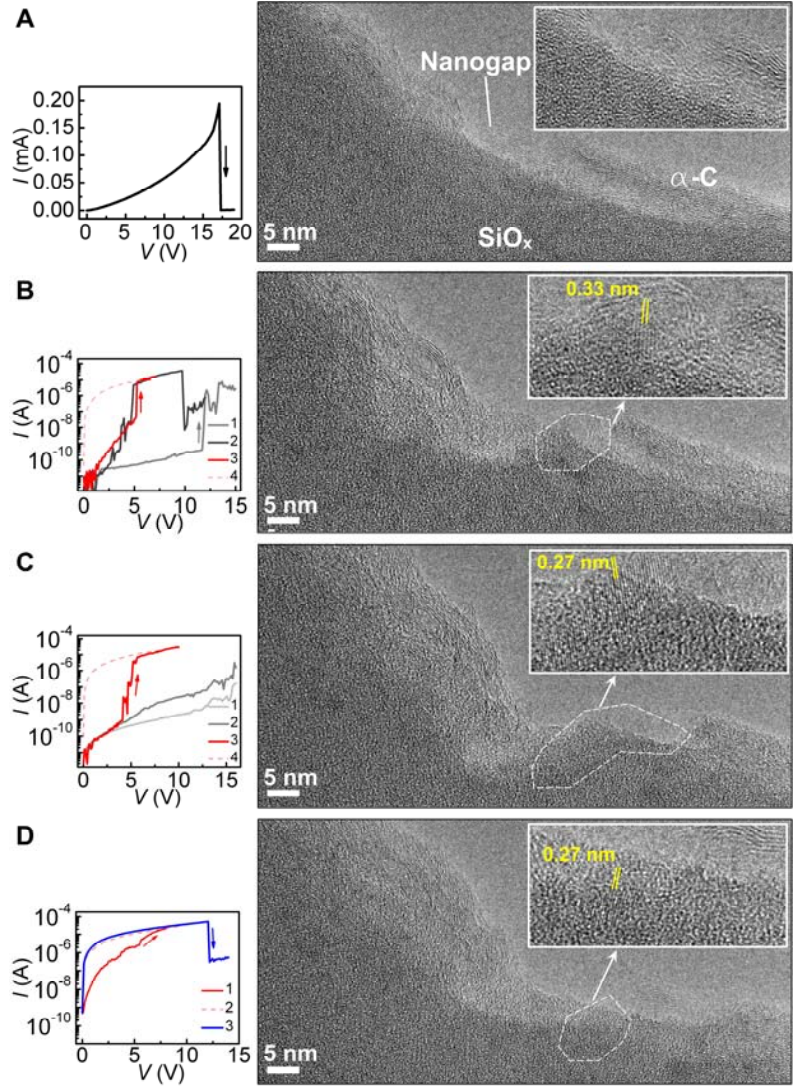

**Fig. 3.** High-resolution TEM images of the nanogap region (right panels) taken immediately after the electrical characterizations (left panels). (**A**) The electrical breakdown *I-V* curve in the α-C layer and the formed nanogap. The inset shows the enlarged nanogap region which shows only amorphous features. (**B**) The electroforming process in SiO$_x$ with the state set to ON. The inset is an enlarged image of the circled region, showing nanocrystalline features. (**C**) A re-electroforming process in SiO$_x$ after the imaging in (B), with the state set to ON. The inset shows the enlarged image of the circled region, showing the growth of the nanocrystal. (**D**) A reset process after the imaging performed in (C). The inset shows the enlarged image of the circled region, showing the shrinkage of size in the nanocrystal. The numbers in all the left-panel plots indicate the voltage-sweep orders.

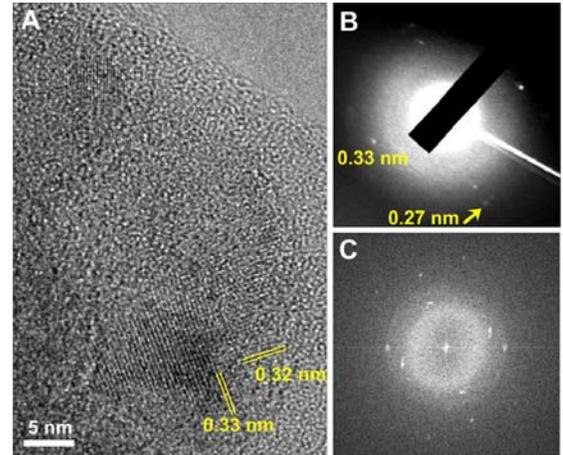

**Fig. 4.** TEM and diffraction patterns from a different device than that shown in Fig. 3. (**A**) High-resolution TEM image of the nanogap region from another SiO$_x$ switching device. (**B**) Selected-area diffraction pattern of the region shown in (A). (**C**) A fast Fourier transformed micrograph of the image shown in (A).

The nanocrystalline structures were persistently observed only at the nanogap region in the switching devices, indicating their correlation with the electroforming and switching processes. Selected area electron diffraction shows that the crystalline structures are consistent with Si nanocrystals (NCs) and not silicon carbide (Fig. S2). As SiO$_x$ is the only source containing the Si element, the formation of the Si NCs shows that the energetically viable SiO$_x$→Si process is associated with the electroforming process. The *in situ* recording of this process also excludes the possibility of processing-induced formation during the filament isolation as was involved in *ex situ* imaging. This TEM imaging process was studied over four different devices (Figs. S2 and S3), all yielding similar information.

As the ON state shows largely metallic conduction (*15*), questions arise regarding the composition of the conductor since conventional silicon is semiconducting. Lattice-spacing measurements and electron diffraction patterns from the Si NCs show evidence of structural deviation from the conventional diamond cubic Si-I phase. As shown in Fig. 4, the intersected lattice spacing of 0.33 and 0.32 nm with a nearly perpendicular angle is the feature of the Si-III phase (*16*), which is semi-metallic (*17*) and often associated with other conducting phases (*18*). This indication can also be seen in Fig. 3; while the lattice spacing of 0.27 nm can be assigned to Si-I(200), Si-III(211) or Si-XII(11-2) faces, that of 0.33 nm (Fig. 3B) can only be associated with Si-III/Si-XII phases (*16*). In particular, as Si-I(200) is the forbidden plane for diffraction (*19*), the strong signal of 0.27 nm$^{-1}$ in the diffraction patterns is further indication of the conducting Si-III/Si-XII phases (*20*) (Figs. S2 and S4). While these phase transitions in Si are typically induced by pressurization (*18*), here it is possible that the high electric field attained at the nanogap is inducing these phases. The stability of the formed silicon phases at ambient environment (*17, 18*) also accounts for the electronically nonvolatile property in SiO$_x$ resistive switching memory. Remarkably, the suppression of the electroforming and switching in SiO$_x$ at low temperature (*9, 15*) coincides with the fact that the Si-III phase cannot be formed at liquid-nitrogen temperature (*21*). Furthermore, the semi-metallic NCs suggest a rationale for failure to induce gating in three-terminal embodiments of these devices (*22*), which was originally proposed to be carbon switching (*22*) and later clarified as SiO$_x$-derived switching (*9, 12*).

From the devices studied, the filament in SiO$_x$ is not in the form of a continuous single crystal across the nanogap, as seen in TiO$_x$ switching systems (*8*). Instead, discrete Si NCs form across the nanogap. This feature indicates a possible transition region between the Si NCs that could favor the switching process. The growth of the Si NCs (Fig. 3C) indicates the general Si-rich nature along the nanogap in the electroformed device. This Si enrichment is also evidently correlated with the indention and shrinkage of the volume at the nanogap region, likely a result of oxygen outgassing (*14*). The shrinkage of the Si NCs with respect to the conductance drop (Fig. 3D) indicates the possible amorphization process. This is consistent with the thermally induced amorphization observed in the metallic Si phases (*18, 20*), as the reset process in unipolar resistive switching is largely thermally stress-driven (*2, 9, 23*). In particular, resistance increase is associated with the amorphization process (*18*). This provides a possible scenario for the filament rupture in the reset process.

We further take the electron-beam impact into consideration during the mechanistic interpretations. As mentioned above, the electron beam from the TEM system tends to degrade the conduction and switching state in SiO$_x$. This does not indicate a charge-based mechanism though, as that was ruled out in the x-ray irradiation experiment and high temperature stability measurements (*9*). In fact, knock-on structural change in Si can be readily induced by an electron beam at the imaging energy (200 KeV) (*24*), and the amorphization process can be induced (*16, 25*). This accounts for the switching degradation after beam exposure (Fig. 3B) as structural changes along the entire filament are induced. While this is a further indication of structural change-induced conductance switching in the Si filament, it also implies that the structural transition needed for the switching can be subtle.

In summary, the study here provides an overall picture of the intrinsic resistive switching in SiO$_x$. The electroforming is through the SiO$_x$→Si process with the semi-metallic Si state identified. The switching is indicated to be through the transition between the semi-metallic and amorphous Si forms. It also provides a general overview of electrical breakdown in silicon oxides. The degradation of the resistive switching state to a non-switchable metallic state (hard breakdown (*26*)) in SiO$_x$ is likely to be associated with the further aggregation of the metallic Si forms. The method described here can also be applied to other resistive switching materials for mechanistic investigation.

## Supporting Information

The multi-stage $SiO_x$ thin-wall structure as shown in Fig. 2 in the main article is fabricated from a silicon wafer (thickness ~ 500 μm) capped with 2 μm thermal $SiO_x$ ($x \sim 2$) on top. It includes the following steps (Fig. S1):

1. Cr sacrificial mask (~ 400 nm thick) is defined by photolithography. Reactive ion etching (RIE) is then used to define the Si stage as shown in Fig. 2B. The etching depth for the Si stage is ~ 12 μm. $CHF_3/O_2$ and $SF_6/Ar/O_2$ recipes are used for $SiO_x$ and Si etching, respectively. The Cr mask is then removed by Cr etchant (CEP-200).

2. At 900 ºC, a layer of α-C is grown on top of the defined Si stage by chemical vapor deposition (CVD) method using $C_2H_2/H_2$ (50/150 sccm) *(1)*. The α-C layer is ~ 20 nm thick with its resistivity ~1-2 KΩ/□. Note that low resistivity is desirable so that the resistance of the α-C electrode is comparatively lower than that of the ON state in $SiO_x$.

3. RIE ($O_2$) is performed for selected-area removal of the α-C layer, with photoresist (S1813) as sacrificial mask defined by photolithography.

4. Electrodes (Ti/Au = 5/50 nm, for wire bonding) is then defined by photolithography. Note the selected-area removal of α-C layer in step 3 is to ensure a direct $SiO_x$/Ti/Au contact for good adhesion during wire bonding.

5. Pt wires that connect the α-C layer and the Au electrodes are defined by electron-beam lithography (EBL). The two Pt wires are separated by ~ 3 μm atop the α-C region.

6. Narrow Cr stripe (~ 140 nm wide, 50 nm thick) between the Pt wires atop the α-C layer is defined by EBL. It serves as the sacrificial mask for $SiO_x$ etching during the definition of the thin-wall $SiO_x$ structure as shown in Fig. 2C.

7. A second Cr mask (~40 nm) is defined by photolithography to protect the Au electrodes during $SiO_x$ etching. Note the Pt wires need no protection as they are resilient to the $SiO_x$-etching recipe.

8. RIE is used to define the $SiO_x$ thin-wall structure with a height of ~ 500 nm (Fig. 2C). Because of undercut, the 140 nm-wide Cr mask eventually results in the width of the $SiO_x$ thin-wall structure ~ 100 nm (Fig. 2D). And the Cr mask is then removed.

The fabricated structure above is then sliced into slim piece (300 μm ×3 mm, Fig. 2A) by a dicing saw. With the height of the $SiO_x$ thin-wall structure ~ 500 nm (Fig. 2C), the width and height of the Si stage ~ 20 μm and 10 μm, respectively (Fig. 2B), the width of the device ~ 300 μm, the fabricated device is expected to have an angle-mismatch tolerance of ~ 0.5 degree for the electron beam. The sliced device is then vertically mounted on a home-built TEM stage that is capable of electrical input, and electrically connected through wire bonding. An Agilent B1500 semiconductor parameter analyzer is used for the electrical characterizations. The imaging is carried out on a JEM-2100F TEM system with the beam energy at 200 KeV.

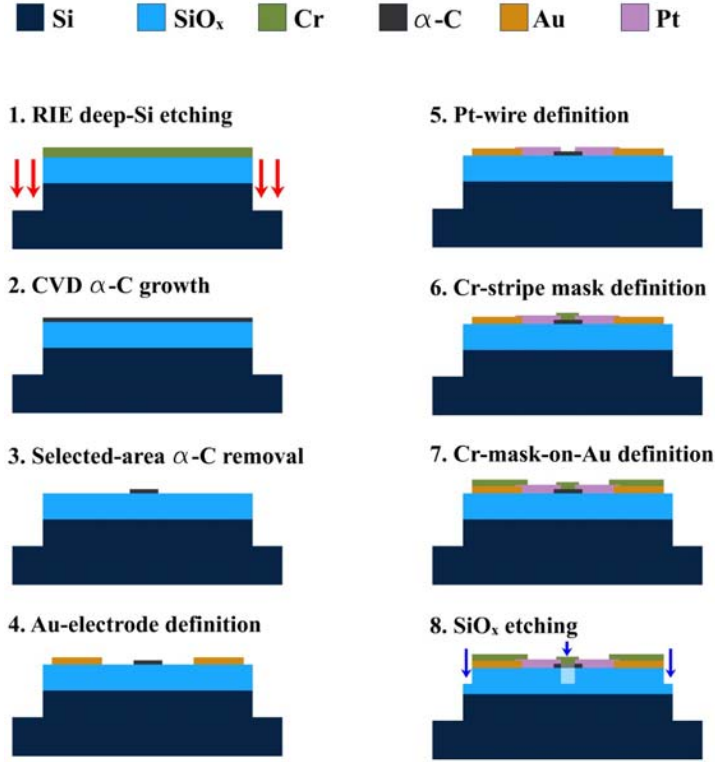

**Fig. S1.** Fabrication flow of the device for the *in situ* TEM imaging device.

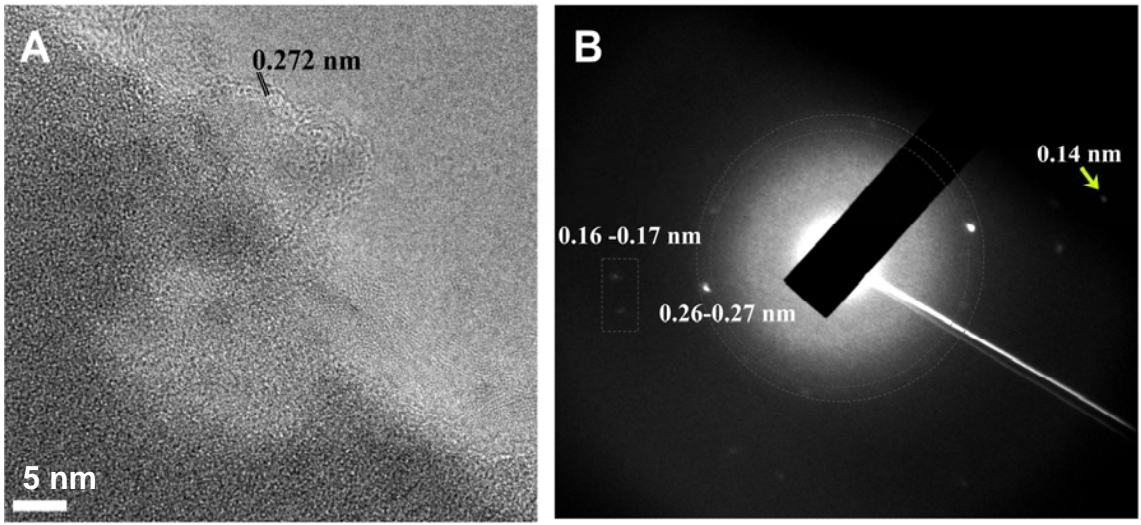

**Fig. S2**. **(A)** High-resolution TEM image of a third switching nanogap (the first and second are in Figs. 3 and 4, respectively) showing the corresponding selected-area diffraction pattern for the device. **(B)** The lattice spacing of 0.27 nm and 0.16 nm and 0.14 nm can be assigned to Si-I (200), (311) and (400) planes, respectively. However, (200) plane is the forbidden plane for diffraction in diamond cubic structure *(2)*. Instead, it is more reasonable to assign the strong signal at 0.27 nm$^{-1}$ to Si-III/Si-XII phases *(3)*. The lattice spacing of 0.27 nm and 0.16 nm and 0.14 nm correspond to Si-III (211), (400) and (422) planes *(4, 5)*. More evidence is shown in Fig. S3 below. In additional, the diffraction pattern here is different from those in the SiC forms *(5-7)* and crystalline silica *(5),* ruling out the possibility of all those forms.

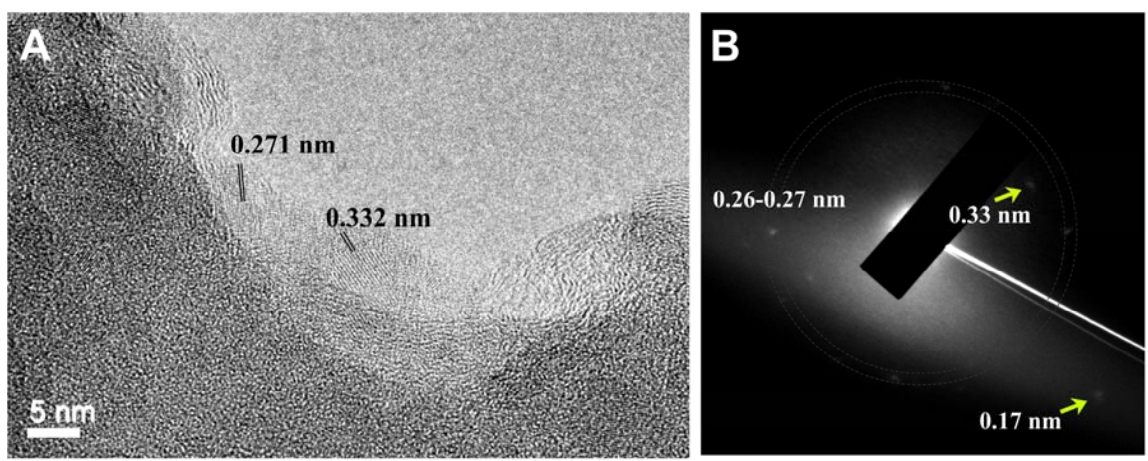

**Fig. S3. (A)** High-resolution TEM image of a fourth switching nanogap and **(B)** the corresponding selected-area diffraction pattern. The diffraction pattern here shows similar feature as that in Fig. S2(B). The signal at 0.33 nm$^{-1}$ can only be assigned to Si-III (211) or its closely-related phase Si-XII (11-1) *(4)*. This lattice spacing along with the absence of 0.31 nm$^{-1}$ signal (corresponding to Si-I(111)) further indicates the Si-III/Si-XII phases.